\documentclass[a4paper,11pt,slantedGreek]{article}
\usepackage[british]{babel}             
\usepackage{pos}
\renewcommand{\logo}{\relax}
\usepackage{graphicx}	
\DeclareRobustCommand{\VAN}[3]{#2}
\let\VANthebibliography\thebibliography
\def\thebibliography{\DeclareRobustCommand{\VAN}[3]{##3}\VANthebibliography}
\hypersetup{pdfauthor={R. E. Spencer et al.},
               pdftitle={Minor Flares on Cygnus X-3 -- VLBI Prospects},       
               bookmarksnumbered=true}

\title{Minor Flares on Cygnus X-3 -- VLBI Prospects}

\title{Minor Flares on Cygnus X-3 -- VLBI Prospects}

\author*[a]{Ralph E.\ Spencer}
\author[a]{Justin D.\ Bray}
\author[b]{David A.\ Green}
\author[a]{Michael A.\ Garrett}

\affiliation[a]{Jodrell Bank Centre for Astrophysics, Dept.\ of Physics and Astronomy,\\ The University of Manchester, Oxford Rd., Manchester M13 9PL, UK}

\affiliation[b]{Cavendish Laboratory,\\ 19 J. J. Thomson Ave., Cambridge, CB3 0HE, UK}

\emailAdd{ralph.spencer@manchester.ac.uk}
\emailAdd{justin.bray@manchester.ac.uk}
\emailAdd{d.a.green@mrao.cam.ac.uk}
\emailAdd{michael.garrett@manchester.ac.uk}

\abstract{
The cm-wavelength radio flares on Cygnus X-3 have been studied for many years. Our recent paper~\citep{Spencer2022} looked again at the minor flares (flux density $S$ of a few 100~mJy) and compared their properties with those of a sample of major flares ($S > 1$~Jy). We find that the minor flares have rise times and duration of $\sim 1$ hour, as opposed to $\sim$ days for the major flares. Minor flares show more rapid expansion of the synchrotron radiation emitting material than in the strong flares. They also appear closer to the binary, whereas the large flares form a more developed jet, i.e.\ the jets formed in minor flares are short and wide, those in major flares are long and thin. We used the results of Fender \& Bright~\citep{FenderBright2019} to calculate the magnetic field and expansion velocity as a fraction $\beta$ of the speed of light under minimum energy conditions when the source is optically thick for samples of minor and major flares. The minimum power in the source was found using the rise time of the flares. The minor flares have lower minimum power but have larger velocities and energy densities than the major flares. Minor flares can occur while a major flare is in progress, suggesting an indirect coupling between them. The spectral evolution of the minor flares can be explained by either an expanding synchrotron source or a shock model. Further investigation requires high resolution VLBI observations at the 1~mas level if we wish to understand the development of the source. The problem is that Cygnus X-3 is strongly scattered by the interstellar medium so high frequencies in the several 10s of GHz are required for the resolution needed. The minor flares are rapid and require high cadence observations to follow the flux density behaviour and hence only short snapshot VLBI observations can capture the structure. Large numbers of telescopes are required which is a problem at the highest frequencies. We discuss the VLBI possibilities and trade-offs for this awkward object.
\vfill
}

\FullConference{%
  15th European VLBI Network Mini-Symposium and Users' Meeting (EVN2022)\\
  11--15 July 2022\\
  University College Cork, Ireland
}

\begin{document}


\maketitle

\section{Introduction}\label{sec:intro}

Cygnus X-3 is famous for its intense radio outbursts when it briefly becomes one of the brightest objects in the radio sky. The first flaring episode to be extensively studied occurred in 1972~\citep{Gregory1972a}, when the first author of this paper was closely involved with the first interferometric observations~\citep{Anderson1972}. It is an X-ray binary system with a $2.4^{+2.1}_{-1.1}$~M$_{\odot}$ compact object with a high mass Wolf--Rayet companion~\citep{Kerkwijk1996, Zdziarski2013}. The high activity of the object suggests that the compact object is a black hole, though the mass function is uncertain due to high interstellar absorption in the optical band. It has a short binary period (4.8~h) and hence has a close binary orbit ($\sim 3 \times 10^{11}$~cm) with the compact object embedded in an intense stellar wind. The object lies in the Galactic plane at a distance of $7.4 \pm 1.1$~kpc~\citep{McCollough2016}. The presence of radio jets during flaring leads to the object being designated as a Microquasar.

Major radio flares at cm wavelengths occur a few times per year. They rise rapidly over $\sim 1$~day and then decay over a few days, reaching levels of 10--20~Jy~\citep{Gregory1972a, Johnston1986, Waltman1995, Fender1997}. Waltman et~al.~\citep{Waltman1994} found three phases of emission: minor flaring during periods of quiescence at levels of a few 100~mJy, quenching where the flux density at 8~GHz drops below 30~mJy, and major flaring. Quenching occurs for several days before major flares~\citep{Waltman1996} and the hard X-ray flux also drops~\citep{McCollough1999}, suggesting a close relationship between the accretion disk and the ejection of plasmons in a radio jet. 

Comparison with X-rays~\citep{Szostek2008} shows that the radio vs X-ray luminosity plot has a reversed `h' shape. As the X-ray flux increases the radio moves from quiescence to minor flaring, and back to quiescence if the X-ray flux decreases, but can also move to the quenched state if the X-ray flux increases preceding a major radio flare. After the flare the source reverts to a lower X-ray state and minor flaring can occur again. Minor flares have peak flux densities between $\sim 0.1$ and 1~Jy~\citep{Waltman1996}, rise in $<1$~hour, have a duration of 1--3~hours, and have been described recently~\citep{Egron2021, Spencer2022}.

MERLIN, VLA and VLBI observations during major flaring have revealed the presence of relativistic jets with two sided~\citep{Marti2001, MillerJones2004} and single sided ejections~\citep{Mioduszewski2001, Tudose2010}. Expansion velocities vary from $0.3c$~\citep{Spencer1986} to $0.81c$~\citep{Mioduszewski2001}. Apparently superluminal velocities and complex changes in the structure suggest that the jets lie close to the line of sight~\citep{Mioduszewski2001, MillerJones2004, Tudose2010}. Several minor flares were found by Ogley et~al.~\citep{Ogley2001} in observations with MERLIN at 5~GHz , triggered by a major flare.

Few high resolution VLBI observations of minor flares exist. Egron et~al.~\citep{Egron2017} were able to follow the evolution of a minor flare event in VLBI observations triggered by a major flare. They fitted a Gaussian model to the visibilities during the minor flare to find it expanded from 0.6 to 0.9 mas in 4 hours. Newell et~al.~\citep{Newell1998} found  an apparent expansion velocity of $2$--$3c$. The apparent brightness temperature for a source which rises in a short time can be high, suggesting either high inverse Compton losses or Doppler boosting of the radiation, and this is discussed in Section~\ref{sec:comparison}.

Further observations at scales of 1~mas or less are required to understand the evolution of minor flares, in particular to discover the relationship between major and minor flares -- why are they so short in duration compared with flares which are perhaps an order of magnitude stronger? A major issue is the presence of strong interstellar scattering and the compromises that are necessary to mitigate it.

Spencer et~al.~\citep{Spencer2022} compared examples of major and minor flares, and applied the formulae of Fender \& Bright~\citep{FenderBright2019} to find physical parameters assuming self-absorption and minimum energy in optically thick flares. Comparisons between major and minor flares are made in Section~\ref{sec:comparison}, followed by discussion of the VLBI issues in Section~\ref{sec:VLBI} and conclusions in Section~\ref{sec:conclusions}.

\section{Major and Minor Flares}\label{sec:comparison}

\subsection{Comparison}

Spencer et~al.~\citep{Spencer2022} discussed 12 minor and 12 major flares, and calculated the brightness temperature $T_{\rm est}$, magnetic field $B$, total energy $E$ and expansion velocity as a fraction of the speed of light $\beta$ using the single frequency formula of Fender \& Bright~\citep{FenderBright2019}. The total power $P=E/t$ was also calculated, where $t$ is the rise time of the flare. Table \ref{tab:calculations} shows the average properties of the major and minor flares.

\begin{table*}
\centering
\begin{tabular}{lcc}\hline
Parameter              & Major Flares & Minor Flares \\\hline
Flux density (Jy)      & $\sim 10$ & $\sim 0.3$ \\
Rate                   & 2--3 year$^{-1}$ & 2--3 month$^{-1}$ \\
Duration (h)           & $\sim 50$ & $\sim 2$    \\
Rise time (h)          & $\sim 24$ & $\sim 1$ \\
Expansion velocity ($c$) & $0.033\pm0.007$      &  $0.13\pm0.01$    \\
Minimum energy (erg)   & $(2.5\pm0.4) \times 10^{41}$ & $(2.8\pm0.8) \times 10^{39}$ \\
Magnetic field (gauss) & $0.58\pm0.05$   &  $1.2\pm0.2$ \\
Power (erg s$^{-1}$)   & $(5.0\pm1.6) \times 10^{36}$ & $(8.5\pm1.9) \times 10^{35}$ \\
Size (au)              & $\sim 100$ & $\sim 3$ \\\hline 
\end{tabular}
\caption{Average parameters for major and minor flares from Spencer et al.~\citep{Spencer2022}.  Expansion velocity, energy, magnetic field and power are calculated assuming minimum energy and self-absorption.}\label{tab:calculations}
\end{table*}

Major flares are much longer in duration than minor flares, have lower radial expansion velocities, higher minimum total energies but similar magnetic fields. The minimum power required to give the total energy in the rise time of the flares is much less than the Eddington limit. If the power required approached the Eddington limit then the source would be a long way from equipartition, with consequences for the generation of the synchrotron radiation emitting electrons. As discussed in Spencer et~al.~\citep{Spencer2022} the minor flares can be well fitted by an expanding synchrotron emitting plasmon, with losses dominated by adiabatic losses. However the higher frequencies can be suppressed by inverse Compton losses in the radiation field of the Wolf--Rayet star, leading to gamma-ray emission. Evidence for losses affecting 15~GHz observations were found~\citep{Spencer2022} for a flare on 7 May 1995. If the brightness temperatures are high enough then self-Compton losses on the synchrotron radiation field could also become important resulting in strong non-thermal X-ray emission.

\subsection{Brightness Temperature}

Inverse Compton losses are expected to be significant in synchrotron sources if brightness temperatures exceed $\sim 5\times10^{11}$~K \citep[see e.g.][]{Readhead1994}. However observed  brightness temperatures in some variable extragalactic sources are much higher than this, from VLBI angular size measurements~\citep{Kovalev2005} and more particularly from variability studies \citep[e.g.\ Liodakis et~al.][]{Liodakis2017}. The emission from these sources is Doppler boosted, but some losses due to inverse self-Compton radiation are expected. To investigate this for Cygnus X-3 we calculated the observed brightness temperature and compared with those calculated previously~\citep{Spencer2022}. 

The Doppler factor $\delta$ for a relativistic source moving at a bulk velocity $\beta_{\rm b}$ towards the observer at an angle $i$ to the line of sight is
\begin{equation}
    \delta=\frac{1}{\Gamma (1-\beta_{\rm b} \cos i)},
\end{equation}
where $\Gamma$ is the Lorentz factor. This results in an observed brightness temperature, estimated from the timescale of variability and causality argument of 
\begin{equation}
    T_{\rm obs}=\delta^3 T_{\rm int},
\end{equation}
where $T_{\rm int}$ is the intrinsic brightness temperature in the rest frame of the source. A derivation of this commonly used formula~\citep{Readhead1994} is given in the appendix of Liodakis et~al.~\citep{Liodakis2017}. The formula is also used in the talk by Jeffrey Hodgson, these proceedings. We distinguish the bulk velocity $\beta_{\rm b}$ from the radial expansion velocity of a spherical source assumed in the Fender \& Bright~\citep{FenderBright2019} calculation. Doppler boosting or de-boosting  results in a modified estimate of brightness temperature ($T_{\rm est}$) from the assumption of minimum energy and self-absorption, $T_{\rm est}=\delta^{2/17} T_{\rm int}$ as shown by Fender \& Bright. From equation 2 we therefore have
\begin{equation}
      T_{\rm obs}=\delta^{49/17} T_{\rm est}.
\end{equation}
From the Rayleigh--Jeans law the observed brightness temperature for a source expanding radially at a velocity $\beta c$ over a timescale $t$ and reaching a peak flaring flux density $F$ is 
\begin{equation}
     T_{\rm obs}= \frac{ F D^2}{2\pi k (\beta t \nu)^2},
\end{equation}
where $D$ is the distance, $\nu$ the frequency and $k$ the Boltzmann constant. A causality limit is given by $\beta=1$.
 
The observed brightness temperatures for the flares described in~\citep{Spencer2022} vary from $6\times 10^{10}$ to $6\times 10^{11}$~K for the major flares and $5\times 10^{10}$ to $5\times 10^{11}$~K for the minor flares, i.e.\ essentially the same with the lower expansion velocity and higher flux density of the major flares being compensated by the much longer rise time $t$. Table~\ref{tab:temperature} shows the average values for the observed temperature $T_{\rm obs}$ from equation 4, the estimated temperature $T_{\rm est}$ and the average value of $\delta$ from equation 3. 
 
\begin{table*}
\centering
\begin{tabular}{lcc}\hline
Parameter      & Major Flares & Minor Flares \\\hline
$T_{\rm obs}$  (K) & $(2.7\pm0.5) \times 10^{11}$ & $(1.7\pm0.4) \times10^{11}$ \\
$T_{\rm est}$  (K) & $(6.75\pm0.06) \times 10^{10}$ & $(5.3\pm0.1) \times10^{10}$\\
$\delta$           & $1.6\pm0.1$  & $1.4\pm0.1$   \\\hline
\end{tabular}
\caption{Brightness temperatures and Doppler factors for major and minor flares.}\label{tab:temperature}
\end{table*}

The observed temperatures are below the Compton limit, and the comparison with estimated temperatures from minimum energy yield relatively low Doppler factors which are similar for both major and minor flares. Note that brightness temperatures at the causal limit are much lower. Self-Compton losses are therefore expected to be low in these flares, though as noted  in Section~\ref{sec:comparison} inverse Compton losses in the optical radiation field of the Wolf-Rayet star can affect minor flares. Very early stages of major flares could also be affected if the radio emitting plasmons are close to  the binary but then move further out.

If the observed bulk velocity is $\sim 0.3c$ then the jet has to be close to the line of sight (less than $10^{\circ}$ if $\beta=0.33$). The complex structure changes observed by Tudose et~al.~\citep{Tudose2010} suggest this is true at least for some major flares. High resolution VLBI is required to confirm this for minor flares and to confirm the velocity estimates.
 
\section{Prospects for VLBI}\label{sec:VLBI}

There are two main difficulties for VLBI imaging observations of minor flares on Cygnus X-3: the broadening of the angular size by interstellar scattering and rapid variability.

Cygnus X-3 lies in a particularly complex part of the Galaxy and suffers severe interstellar scattering~\citep{Wilkinson1994}. The angular size $\theta$ in mas at frequency $\nu$ in GHz is broadened by~\citep{Mioduszewski2001}
\begin{equation}
    \theta=\frac{448~\textrm{mas}}{(\nu / \textrm{1~GHz})^{2.09}}.
\end{equation}
If resolutions of $\sim 1$~mas or less are required then frequencies >18 GHz are needed. The scattering size at 23~GHz is 0.63~mas, and a source at 7.4~kpc will reach this diameter if expanding at $0.13c$ after $\sim 2.5$ hours, close to the typical duration of a minor flare. It should therefore be possible to detect the radial expansion of a minor flare and also its bulk motion if moving at $0.3c$.

\begin{figure}
\centering
\includegraphics[width=0.9\textwidth]{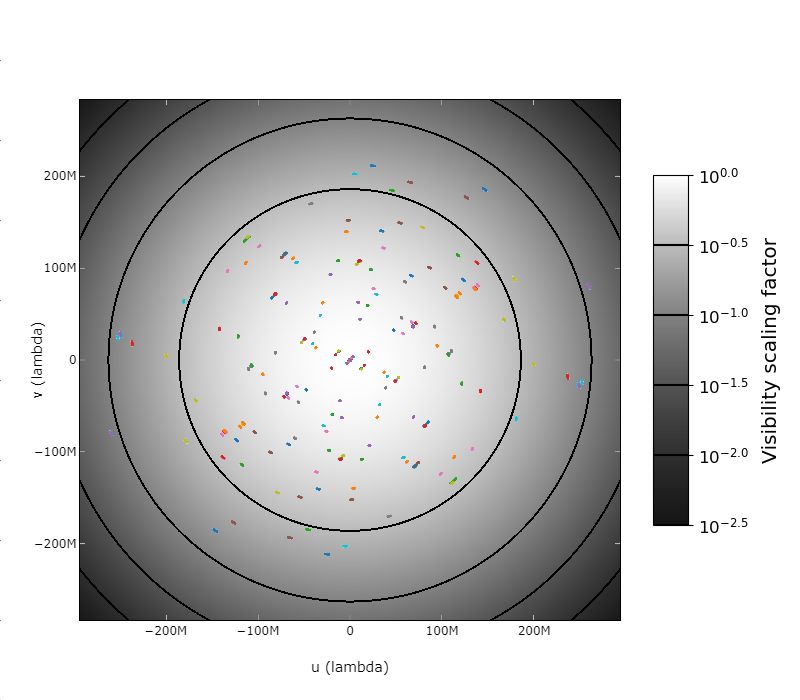}
\caption{Aperture plane coverage for a snapshot observation with EVN at 23~GHz. The colours represent baselines between individual telescopes. The effect of scattering on visibility is shown as a grey scale, with the level of the contours indicated on the side bar. The effect of scattering is also shown in Figure \ref{fig:scatter2}.}\label{fig:scatter}
\end{figure}

Since the source is changing rapidly, simple Earth rotation synthesis is not possible and so snapshot observations are required. Luckily modern VLBI arrays are quite sensitive and good signal to noise observations can be made in integration times of a minute or less. As many telescopes as possible are needed to ensure adequate aperture plane ({\it{u,v}}) coverage. Figure \ref{fig:scatter} shows the {\it{u,v}} coverage for a 6-minute snapshot using the EVN at 23.1 GHz. Also superimposed in grey-scale is the expected visibility factor assuming the scattering results in a Gaussian single source structure. The {\it{u,v}} values are in wavelengths for an observation at 20:21 GST with 15 telescopes (including the high frequency dishes in e-MERLIN and also 2 Russian telescopes -- not currently available). The expected beam is $\sim 1.1 \times 0.8$ mas$^2$ with a noise level of 0.1 mJy (natural weighting). The amplitudes on the longest baselines ($\sim 220$~M$\lambda$) are expected to be reduced by a factor of $\sim 5$ giving a final point source image size of 1.2 mas. Note that baselines to the VLBA from EVN telescopes will have much reduced amplitude due to scattering and so simultaneous observations are not worthwhile.

\begin{figure}
\centering
\includegraphics[width=0.9\textwidth]{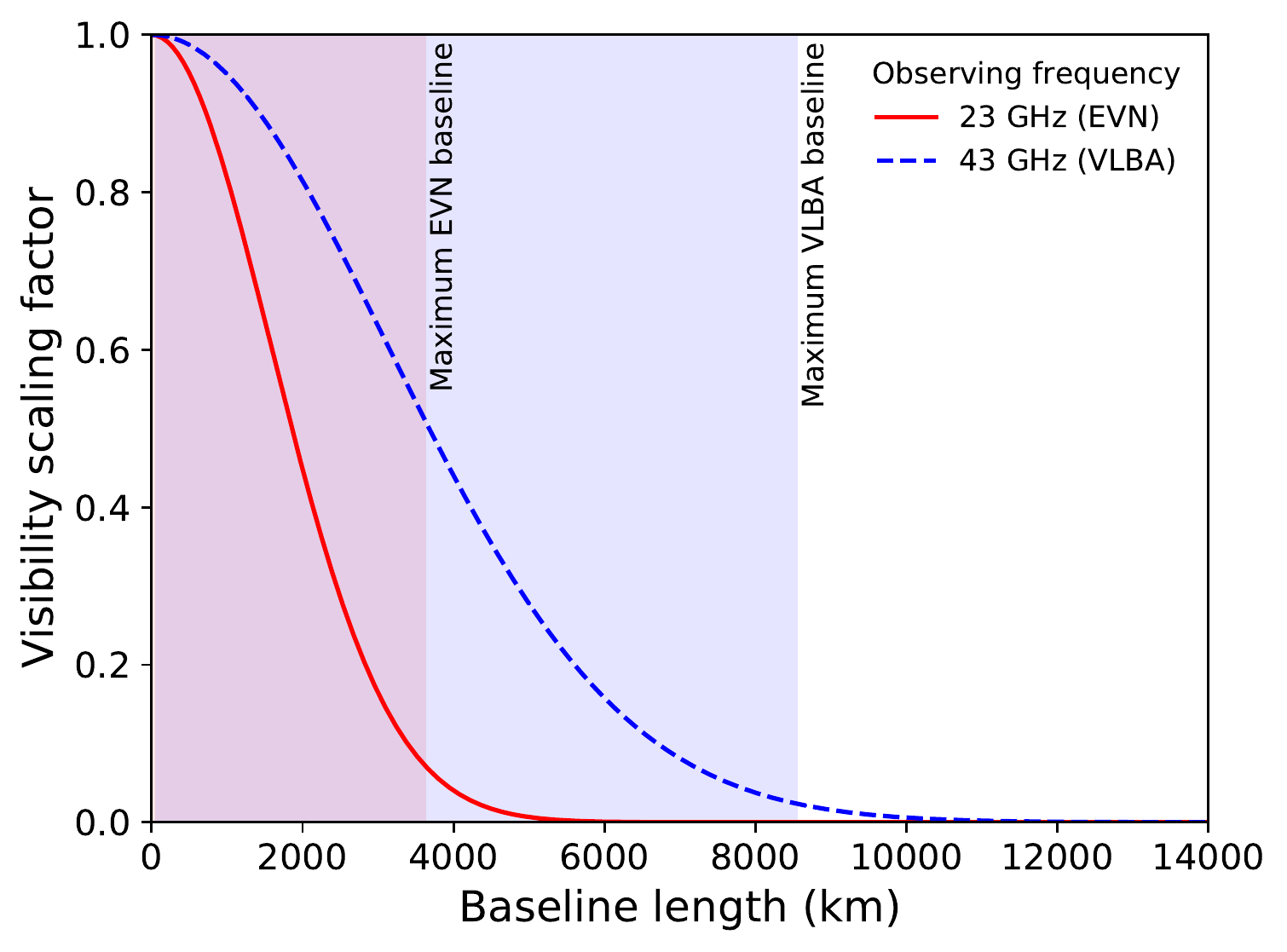}
\caption{Scaling factor for the reduction in visibility of a Gaussian source due to interstellar scattering, for potential observations with the EVN at 23~GHz and the VLBA at 43~GHz.  Filled regions show the range of baseline lengths for each instrument, with the longest baselines labelled.  In both cases, visibilities on the longest baselines will be <10\% of the values that would be observed without scattering.}\label{fig:scatter2}
\end{figure}

The scattering size at 43 GHz is 0.17 mas, and so there is a much improved opportunity to measure the structure changes  of a minor flare.  The EVN has few telescopes operating at this frequency but the VLBA with its 10 telescopes has reasonable {\it{u,v}} coverage with a beam of $\sim 0.6 \times 0.3$ mas$^2$ and 0.2~mJy sensitivity depending on weighting and GST. Figure \ref{fig:scatter2} shows the affect of scattering on visibility for the EVN at 23 GHz and the VLBA at 43 GHz. It should be possible to follow the evolution of a minor flare with the VLBA, though the problem of catching one or more minor flares during a random say 10-hour observation run remains.

\section{Conclusions}\label{sec:conclusions}

Physical parameters of the emitting regions of major and minor radio flares on Cygnus X-3 have been found assuming the minimum energy and sef-absorption condition in the flares. While the total energy and power  in major flares are higher, the magnetic fields responsible for synchrotron emission and estimated brightness temperatures are similar.  The higher expansion velocities found for minor flares are a direct consequence of their different time scales. 
We therefore expect the plasmons or jets responsible for minor flares to be more compact than those in major flares and to expand and evolve more quickly. Brightness temperatures found from variability time scales are similar, and require relatively low Doppler factors, unlike those in extra-galactic variable sources. High resolution VLBI observations are needed to study the evolving structure of minor flares, and to clarify the relationship with major flares. 

Cygnus X-3 evolves on short timescales (< 1 day for major flares and $\sim$1\ hour for minor flares) and therefore multiple  short snap-shot observations are preferred. Luckily the source is quite strong and so good signal/noise can be obtained even for observations of a few minutes.  The strong interstellar scattering in the direction of Cygnus X-3 means that high frequency observations are required, but then fewer telescopes are available and hence give poorer aperture plane coverage. Compromises have to be made. The quenching observed before major outbursts can act as a trigger for VLBI observations as a target of opportunity, but such a trigger is not known for minor flares. One just has to be lucky, and there is therefore a chance that valuable observing time may be wasted.

\acknowledgments
We thank Marcello Giroletti for pointing out the paper by Egron et~al.~\citep{Egron2017}, and a participant at the conference for suggesting we look at the brightness temperature in more detail. RES thanks the directors of JBCA for providing travel funds to attend the conference. We made extensive use of the EVN observation planner (available at \href{https://www.evlbi.org/}{https://www.evlbi.org/}, viewed 16/09/2022).

\bibliographystyle{JHEP}
\bibliography{XRB.bib,new.bib}

\providecommand{\href}[2]{#2}\begingroup\raggedright\begin{thebibliography}{10}

\bibitem{Spencer2022}
R.E.~{Spencer}, M.~{Garrett}, J.D.~{Bray} and D.A.~{Green}, \emph{{Major and
  minor flares on Cygnus X-3 revisited}},
  \href{https://doi.org/10.1093/mnras/stac666}{\emph{\mnras} {\bfseries 512}
  (2022) 2618} [\href{https://arxiv.org/abs/arXiv:2203.05637}{{\ttfamily
  arXiv:2203.05637}}].

\bibitem{FenderBright2019}
R.~{Fender} and J.~{Bright}, \emph{{Synchrotron self-absorption and the minimum
  energy of optically thick radio flares from stellar mass black holes}},
  \href{https://doi.org/10.1093/mnras/stz2000}{\emph{\mnras} {\bfseries 489}
  (2019) 4836} [\href{https://arxiv.org/abs/1907.07463}{{\ttfamily
  1907.07463}}].

\bibitem{Gregory1972a}
P.C.~{Gregory} and P.P.~{Kronberg}, \emph{{Discovery of Giant Radio Outburst
  from Cygnus X-3}}, \href{https://doi.org/10.1038/239440a0}{\emph{\nat}
  {\bfseries 239} (1972) 440}.

\bibitem{Anderson1972}
B.~{Anderson}, R.G.~{Conway}, R.J.~{Davis}, R.J.~{Peckham}, P.J.~{Richards},
  R.E.~{Spencer} et~al., \emph{{Observations at 408 MHz of the Cyg X-3 Radio
  Outburst}}, \href{https://doi.org/10.1038/physci239117a0}{\emph{Nature Phys.\
  Sci.} {\bfseries 239} (1972) 117}.

\bibitem{Kerkwijk1996}
M.H.~{\VAN{Kerkwijk}{Van}{van}}~Kerkwijk, T.R.~{Geballe}, D.L.~{King}, M.~{van
  der Klis} and J.~{van Paradijs}, \emph{{The Wolf-Rayet counterpart of Cygnus
  X-3.}}, {\emph{\aap} {\bfseries 314} (1996) 521}
  [\href{https://arxiv.org/abs/astro-ph/9604100}{{\ttfamily
  astro-ph/9604100}}].

\bibitem{Zdziarski2013}
A.A.~{Zdziarski}, J.~{Mikolajewska} and K.~{Belczynski}, \emph{{Cyg X-3: a
  low-mass black hole or a neutron star.}},
  \href{https://doi.org/10.1093/mnrasl/sls035}{\emph{\mnras} {\bfseries 429}
  (2013) L104} [\href{https://arxiv.org/abs/1208.5455}{{\ttfamily 1208.5455}}].

\bibitem{McCollough2016}
M.L.~{McCollough}, L.~{Corrales} and M.M.~{Dunham}, \emph{{Cygnus X-3: Its
  Little Friend{\textquoteright}s Counterpart, the Distance to Cygnus X-3, and
  Outflows/Jets}},
  \href{https://doi.org/10.3847/2041-8205/830/2/L36}{\emph{\apjl} {\bfseries
  830} (2016) L36} [\href{https://arxiv.org/abs/1610.01923}{{\ttfamily
  1610.01923}}].

\bibitem{Johnston1986}
K.J.~{Johnston}, J.H.~{Spencer}, R.S.~{Simon}, E.B.~{Waltman}, G.G.~{Pooley},
  R.E.~{Spencer} et~al., \emph{{Radio Flux Density Variations of Cygnus X-3}},
  \href{https://doi.org/10.1086/164639}{\emph{\apj} {\bfseries 309} (1986)
  707}.

\bibitem{Waltman1995}
E.B.~{Waltman}, F.D.~{Ghigo}, K.J.~{Johnston}, R.S.~{Foster}, R.L.~{Fiedler}
  and J.H.~{Spencer}, \emph{{The Evolution of Outbursts in Cygnus X-3 at 2.25
  and 8.3 GHz}}, \href{https://doi.org/10.1086/117518}{\emph{\aj} {\bfseries
  110} (1995) 290}.

\bibitem{Fender1997}
R.P.~{Fender}, S.J.~{Bell Burnell}, E.B.~{Waltman}, G.G.~{Pooley}, F.D.~{Ghigo}
  and R.S.~{Foster}, \emph{{Cygnus X-3 in outburst: quenched radio emission,
  radiation losses and variable local opacity}},
  \href{https://doi.org/10.1093/mnras/288.4.849}{\emph{\mnras} {\bfseries 288}
  (1997) 849} [\href{https://arxiv.org/abs/astro-ph/9612125}{{\ttfamily
  astro-ph/9612125}}].

\bibitem{Waltman1994}
E.B.~{Waltman}, R.L.~{Fiedler}, K.J.~{Johnston} and F.D.~{Ghigo}, \emph{{The
  Quiescent Level of Cygnus X-3 at 2.25 and 8.3 GHz: 1988-1992}},
  \href{https://doi.org/10.1086/117056}{\emph{\aj} {\bfseries 108} (1994) 179}.

\bibitem{Waltman1996}
E.B.~{Waltman}, R.S.~{Foster}, G.G.~{Pooley}, R.P.~{Fender} and F.D.~{Ghigo},
  \emph{{Quenched Radio Emission in Cygnus X-3}},
  \href{https://doi.org/10.1086/118213}{\emph{\aj} {\bfseries 112} (1996)
  2690}.

\bibitem{McCollough1999}
M.L.~{McCollough}, C.R.~{Robinson}, S.N.~{Zhang}, B.A.~{Harmon},
  R.M.~{Hjellming}, E.B.~{Waltman} et~al., \emph{{Discovery of Correlated
  Behavior between the Hard X-Ray and the Radio Bands in Cygnus X-3}},
  \href{https://doi.org/10.1086/307241}{\emph{\apj} {\bfseries 517} (1999) 951}
  [\href{https://arxiv.org/abs/astro-ph/9810212}{{\ttfamily
  astro-ph/9810212}}].

\bibitem{Szostek2008}
A.~{Szostek}, A.A.~{Zdziarski} and M.L.~{McCollough}, \emph{{A classification
  of the X-ray and radio states of Cyg X-3 and their long-term correlations}},
  \href{https://doi.org/10.1111/j.1365-2966.2008.13479.x}{\emph{\mnras}
  {\bfseries 388} (2008) 1001}
  [\href{https://arxiv.org/abs/0803.2217}{{\ttfamily 0803.2217}}].

\bibitem{Egron2021}
E.~{Egron}, A.~{Pellizzoni}, S.~{Righini}, M.~{Giroletti}, K.~{Koljonen},
  K.~{Pottschmidt} et~al., \emph{{Investigating the Mini and Giant Radio Flare
  Episodes of Cygnus X-3}},
  \href{https://doi.org/10.3847/1538-4357/abc5b1}{\emph{\apj} {\bfseries 906}
  (2021) 10} [\href{https://arxiv.org/abs/2010.15002}{{\ttfamily 2010.15002}}].

\bibitem{Marti2001}
J.~{Mart{\'\i}}, J.M.~{Paredes} and M.~{Peracaula}, \emph{{Development of a
  two-sided relativistic jet in Cygnus X-3}},
  \href{https://doi.org/10.1051/0004-6361:20010907}{\emph{\aap} {\bfseries 375}
  (2001) 476}.

\bibitem{MillerJones2004}
J.C.A.~{Miller-Jones}, K.M.~{Blundell}, M.P.~{Rupen}, A.J.~{Mioduszewski},
  P.~{Duffy} and A.J.~{Beasley}, \emph{{Time-sequenced Multi-Radio Frequency
  Observations of Cygnus X-3 in Flare}},
  \href{https://doi.org/10.1086/379706}{\emph{\apj} {\bfseries 600} (2004) 368}
  [\href{https://arxiv.org/abs/astro-ph/0311277}{{\ttfamily
  astro-ph/0311277}}].

\bibitem{Mioduszewski2001}
A.J.~{Mioduszewski}, M.P.~{Rupen}, R.M.~{Hjellming}, G.G.~{Pooley} and
  E.B.~{Waltman}, \emph{{A One-sided Highly Relativistic Jet from Cygnus X-3}},
  \href{https://doi.org/10.1086/320965}{\emph{\apj} {\bfseries 553} (2001) 766}
  [\href{https://arxiv.org/abs/astro-ph/0102018}{{\ttfamily
  astro-ph/0102018}}].

\bibitem{Tudose2010}
V.~{Tudose}, J.C.A.~{Miller-Jones}, R.P.~{Fender}, Z.~{Paragi}, C.~{Sakari},
  A.~{Szostek} et~al., \emph{{Probing the behaviour of the X-ray binary Cygnus
  X-3 with very long baseline radio interferometry}},
  \href{https://doi.org/10.1111/j.1365-2966.2009.15719.x}{\emph{\mnras}
  {\bfseries 401} (2010) 890}
  [\href{https://arxiv.org/abs/0909.2790}{{\ttfamily 0909.2790}}].

\bibitem{Spencer1986}
R.E.~{Spencer}, R.W.~{Swinney}, K.J.~{Johnston} and R.M.~{Hjellming},
  \emph{{The 1983 September Radio Outburst of Cygnus X-3: Relativistic
  Expansion at 0.35c}}, \href{https://doi.org/10.1086/164637}{\emph{\apj}
  {\bfseries 309} (1986) 694}.

\bibitem{Ogley2001}
R.N.~{Ogley}, S.J.~{Bell Burnell}, R.E.~{Spencer}, S.J.~{Newell},
  A.M.~{Stirling} and R.P.~{Fender}, \emph{{Radio flares and plasmon size in
  Cygnus X-3}},
  \href{https://doi.org/10.1046/j.1365-8711.2001.04617.x}{\emph{\mnras}
  {\bfseries 326} (2001) 349}
  [\href{https://arxiv.org/abs/astro-ph/0104469}{{\ttfamily
  astro-ph/0104469}}].

\bibitem{Egron2017}
E.~{Egron}, A.~{Pellizzoni}, M.~{Giroletti}, S.~{Righini}, M.~{Stagni},
  A.~{Orlati} et~al., \emph{{Single-dish and VLBI observations of Cygnus X-3
  during the 2016 giant flare episode}},
  \href{https://doi.org/10.1093/mnras/stx1730}{\emph{\mnras} {\bfseries 471}
  (2017) 2703} [\href{https://arxiv.org/abs/arXiv:1707.03761}{{\ttfamily
  arXiv:1707.03761}}].

\bibitem{Newell1998}
S.J.~{Newell}, M.A.~{Garrett} and R.E.~{Spencer}, \emph{{Apparent superluminal
  expansion in Cygnus X-3}},
  \href{https://doi.org/10.1046/j.1365-8711.1998.01230.x}{\emph{\mnras}
  {\bfseries 293} (1998) L17}.

\bibitem{Readhead1994}
A.C.S.~{Readhead}, \emph{{Equipartition Brightness Temperature and the Inverse
  Compton Catastrophe}}, \href{https://doi.org/10.1086/174038}{\emph{\apj}
  {\bfseries 426} (1994) 51}.

\bibitem{Kovalev2005}
Y.Y.~{Kovalev}, K.I.~{Kellermann}, M.L.~{Lister}, D.C.~{Homan},
  R.C.~{Vermeulen}, M.H.~{Cohen} et~al., \emph{{Sub-Milliarcsecond Imaging of
  Quasars and Active Galactic Nuclei. IV. Fine-Scale Structure}},
  \href{https://doi.org/10.1086/497430}{\emph{\aj} {\bfseries 130} (2005) 2473}
  [\href{https://arxiv.org/abs/astro-ph/0505536}{{\ttfamily
  astro-ph/0505536}}].

\bibitem{Liodakis2017}
I.~{Liodakis}, N.~{Marchili}, E.~{Angelakis}, L.~{Fuhrmann}, I.~{Nestoras},
  I.~{Myserlis} et~al., \emph{{F-GAMMA: variability Doppler factors of blazars
  from multiwavelength monitoring}},
  \href{https://doi.org/10.1093/mnras/stx002}{\emph{\mnras} {\bfseries 466}
  (2017) 4625} [\href{https://arxiv.org/abs/1701.01452}{{\ttfamily
  1701.01452}}].

\bibitem{Wilkinson1994}
P.N.~{Wilkinson}, R.~{Narayan} and R.E.~{Spencer}, \emph{{The Scatter-Broadened
  Image of CYGNUS-X-3}},
  \href{https://doi.org/10.1093/mnras/269.1.67}{\emph{\mnras} {\bfseries 269}
  (1994) 67}.

\end{thebibliography}\endgroup

\end{document}